\documentclass[11pt,a4paper]{article}
\usepackage[utf8]{inputenc}
\usepackage[T1]{fontenc}
\usepackage{lmodern}
\usepackage[margin=1in]{geometry}
\usepackage{amsmath}
\usepackage{graphicx}
\usepackage{booktabs}
\usepackage{hyperref}
\usepackage{url}
\usepackage{natbib}
\usepackage{xcolor}
\usepackage{enumitem}
\usepackage{tabularx}
\usepackage{array}
\usepackage{longtable}
\hypersetup{
  colorlinks=true,
  linkcolor=blue,
  citecolor=blue,
  urlcolor=blue
}
\title{From Theory to Protocol: Executable Frameworks for Creative Emergence and Strategic Foresight}
\author{Shun Fujiyoshi\\
Independent Researcher\\
\url{https://github.com/GhostyAI-HA}}
\date{February 2026}
\begin{document}
\maketitle
\begin{abstract}
Creativity and strategic foresight have been extensively studied through descriptive theories --- Koestler's bisociation (1964), de Bono's lateral thinking (1967), and Ansoff's weak signals (1975) explain \emph{why} creative and strategic insights occur, but offer limited guidance on \emph{how to produce them on demand}. This paper presents two executable protocols that bridge this theory-practice gap: \textbf{GHOSTY COLLIDER}, a 5-step protocol for cross-domain creative emergence through structural de-labeling and collision, and \textbf{PRECOG PROTOCOL}, a 5-step protocol for signal-based strategic foresight with multi-axis timing judgment. We formalize established theories into repeatable, step-by-step procedures with explicit quality criteria, anti-pattern detection, and measurable outputs. We evaluate the protocols through three complementary methods: (1) five detailed case studies across distinct domains, (2) controlled comparisons against standard methods using identical inputs, and (3) a batch experiment across eight random domain pairings (N=8, success rate 87.5\%, failure rate 12.5\%) with one blind evaluation. Preliminary evidence suggests that protocol-driven outputs exhibit greater structural novelty, higher parameter specificity, and qualitatively distinct creative directions compared to outputs from standard methods. The blind evaluation confirmed the direction of author assessments (protocol output scored 74/80 vs.\ brainstorming 49/80). These results, while limited by single-operator execution, indicate that the theory-to-protocol translation preserves and potentially enhances the generative power of the underlying theories. The protocols, updated to version 2 incorporating lessons from failure case analysis, are released as open-access documents under CC BY-NC 4.0 at \url{https://github.com/GhostyAI-HA/ghosty-collider}.
\end{abstract}
\section{Introduction}
The gap between creativity theory and creative practice is well-documented. Sawyer \citeyearpar{sawyer2012} notes that decades of creativity research have produced rich explanatory models but few actionable tools for practitioners. Similarly, in strategic foresight, Rohrbeck et al.\ \citeyearpar{rohrbeck2015} observe that while scanning and scenario methodologies are theoretically mature, their translation into organizational practice remains challenging --- requiring multi-week workshops, trained facilitators, and significant resources.
Recent advances in large language models (LLMs) have created a new context for this gap. Stevenson et al.\ \citeyearpar{stevenson2022} demonstrated that GPT-3 performs competitively on divergent thinking tasks, while Doshi and Hauser \citeyearpar{doshi2024} found that LLM-assisted ideation generates more ideas but may reduce diversity across participants. Gero and Chilton's Metaphoria \citeyearpar{gero2019} showed that algorithmic systems can generate non-obvious metaphorical connections, and their subsequent work \citep{gero2023} examined how AI support affects the social dynamics of creative writing. These findings suggest that LLMs are powerful ideation engines, but that \emph{unstructured} AI-assisted ideation risks producing outputs that converge to genre conventions --- what Doshi and Hauser \citeyearpar{doshi2024} term the ``flattening effect.'' This observation motivates our central question: \textbf{Can structured protocols, grounded in established creativity and foresight theories, produce qualitatively different outputs from unstructured AI-assisted ideation?}
We present two protocols designed to answer this question:
\begin{enumerate}
  \item \textbf{GHOSTY COLLIDER} transforms creativity theories (bisociation, lateral thinking, Geneplore model) into a 5-step procedure: Fragment Harvest $\rightarrow$ Ghost Extraction $\rightarrow$ Collision Matrix $\rightarrow$ Vision Crystallization $\rightarrow$ Reality Bridge. The key innovation is \emph{Ghost Extraction} --- a systematic de-labeling process that reduces concepts to their deep structural elements (verbs, forces, transformations) before collision, increasing the probability of non-obvious cross-domain connections.
  \item \textbf{PRECOG PROTOCOL} transforms strategic foresight methodologies (horizon scanning, weak signals theory, scenario planning) into a 5-step procedure: Signal Map $\rightarrow$ Convergence Analysis $\rightarrow$ Contrarian View $\rightarrow$ Timing Grid $\rightarrow$ Action Window. The key innovations are the \emph{multi-axis Timing Grid} (evaluating market phase, competitive position, organizational readiness, and external windows simultaneously) and \emph{Signal History tracking} for longitudinal foresight.
\end{enumerate}
The protocols share a design philosophy: strip away the scaffolding of workshops, facilitators, and multi-day processes, preserving only the core cognitive operations that produce insight. They are designed to be executed by individuals, with or without AI assistance, in a single session.
We wish to be explicit about the scope of our empirical contribution. The evidence presented here constitutes \textbf{preliminary, proof-of-concept evaluation}: all case studies and controlled comparisons were executed by a single operator (the author) with AI assistance. We present this evidence as initial demonstration of the protocols' structural properties rather than as definitive proof of general effectiveness. Multi-user evaluation across diverse skill levels and contexts is essential future work (Section~\ref{sec:future}).
To evaluate the protocols, we conduct three forms of assessment: (1) multi-domain case studies demonstrating protocol execution across five distinct domains, (2) controlled comparisons where identical inputs are processed with and without the protocols, using standard methods (brainstorming, SWOT analysis) as baselines, and (3) a batch experiment across eight random domain pairings to measure protocol success/failure rates, with one blind evaluation to partially address evaluator bias.
The remainder of this paper is organized as follows: Section~\ref{sec:related} reviews the theoretical foundations. Section~\ref{sec:ghosty} presents GHOSTY COLLIDER. Section~\ref{sec:precog} presents PRECOG PROTOCOL. Section~\ref{sec:integration} describes their integration. Section~\ref{sec:cases} demonstrates both protocols through five case studies. Section~\ref{sec:comparisons} presents controlled comparisons against standard methods, reports protocol failure cases with quantified failure rates, and presents a blind evaluation. Section~\ref{sec:discussion} discusses contributions and limitations. Section~\ref{sec:conclusion} concludes.
\section{Related Work}
\label{sec:related}
\subsection{Creativity Theories}
\textbf{Bisociation Theory.} Koestler \citeyearpar{koestler1964} introduced bisociation as the simultaneous perception of an idea in two self-consistent but habitually incompatible ``matrices of thought.'' Unlike association (connecting ideas within a single frame), bisociation requires the collision of separate frames --- the ``Eureka moment'' occurs when a connection is perceived across frames that were previously considered unrelated. GHOSTY COLLIDER operationalizes this through Ghost Extraction (stripping frames) and Collision Matrix (forcing cross-frame intersection).
\textbf{Lateral Thinking.} De Bono \citeyearpar{debono1967,debono1970} formalized the concept of lateral thinking as a deliberate method for breaking out of dominant patterns. His key insight --- that logical, sequential thinking is insufficient for creative leaps because it stays within established patterns --- motivates GHOSTY COLLIDER's de-labeling step. By removing the label (the pattern), the thinker is forced into a lateral mode.
\textbf{Geneplore Model.} Finke, Ward, and Smith \citeyearpar{finke1992} proposed a two-phase model of creativity: the \emph{generative phase} (producing pre-inventive structures --- ambiguous, incomplete mental representations) and the \emph{exploratory phase} (interpreting, refining, and evaluating these structures). GHOSTY COLLIDER maps directly onto this model: Steps 1--3 (Fragment Harvest through Collision Matrix) constitute the generative phase, producing pre-inventive structures through de-labeling and collision; Steps 4--5 (Vision Crystallization through Reality Bridge) constitute the exploratory phase, interpreting emergent patterns and grounding them in reality.
\textbf{CK Design Theory.} Hatchuel and Weil \citeyearpar{hatchuel2003,hatchuel2009} proposed Concept-Knowledge (CK) theory, which models design as a dynamic interplay between a \emph{Concept space} (propositions that are neither true nor false --- undecided) and a \emph{Knowledge space} (established propositions). Innovation occurs when operators expand the concept space beyond the boundaries of current knowledge. GHOSTY COLLIDER's Ghost Extraction parallels CK theory's concept expansion: by stripping domain-specific knowledge (labels), the protocol forces concepts into undecided territory, enabling novel partitions that would not emerge within a single knowledge frame.
\textbf{Systematic Inventive Thinking (SIT).} Horowitz \citeyearpar{horowitz2001} developed SIT as a structured approach where creative solutions follow identifiable patterns (subtraction, multiplication, division, task unification, attribute dependency change). Unlike GHOSTY COLLIDER, which operates on domain-agnostic structural elements, SIT applies modification operators to the existing form of a product or process. The approaches are complementary: SIT optimizes within a domain, while GHOSTY COLLIDER generates cross-domain visions.
\textbf{Existing Creativity Methods.} Several structured methods exist for creative production. Osborn's \citeyearpar{osborn1953} brainstorming generates ideas through free association but remains within familiar frames. Eberle's \citeyearpar{eberle1996} SCAMPER provides systematic modification operators (Substitute, Combine, Adapt, etc.) but is bounded by the existing form of the object. Altshuller's \citeyearpar{altshuller1946} TRIZ offers powerful inventive principles for engineering problems but requires domain-specific technical contradictions as input. The Design Thinking framework \citep{brown2008} centers the process on user empathy but requires user research as a prerequisite. GHOSTY COLLIDER differs from all of these in its input (raw, heterogeneous fragments from any domain), its process (de-labeling before combination), and its output (a named vision of something that does not yet exist).
\subsection{AI-Augmented Creativity}
The emergence of large language models has created a new landscape for creativity support tools \citep{frich2019}. Recent work demonstrates both the promise and limitations of AI-assisted ideation.
\textbf{LLM Creative Capabilities.} Stevenson et al.\ \citeyearpar{stevenson2022} evaluated GPT-3 on the Alternative Uses Test --- a standard measure of divergent thinking --- and found that it performed comparably to human participants. Subsequent studies \citep{haase2023} confirmed that LLMs can match or exceed the \emph{average} human on standardized creativity tasks, while the most creative individuals still outperform AI. Si et al.\ \citeyearpar{si2024} developed systematic benchmarks for evaluating LLM scientific ideation, finding that LLMs generate novel ideas at competitive rates but with lower feasibility scores than expert-generated ideas.
\textbf{Human-AI Co-Creation Dynamics.} Gero and Chilton \citeyearpar{gero2019} developed Metaphoria, an algorithmic system that generates metaphorical connections from a writer's input, demonstrating that structured AI assistance can produce non-obvious creative associations. Their subsequent work \citep{gero2023} revealed that the \emph{social dynamics} of AI-assisted creative writing differ significantly from pure human collaboration --- a finding that complicates evaluation of AI-augmented creativity tools. Lee et al.\ \citeyearpar{lee2024} surveyed LLM-based ideation tools and proposed the Hourglass Ideation Framework, identifying that LLMs contribute most effectively to idea generation and refinement but remain underutilized for scope specification and multi-idea evaluation.
\textbf{The Flattening Concern.} Doshi and Hauser \citeyearpar{doshi2024} provided the most directly relevant finding for our work. In a large-scale field experiment, they showed that while AI-assisted writers produced more content, the aggregate \emph{diversity} of outputs decreased --- individual improvements came at the cost of collective homogeneity. This ``flattening effect'' motivates the design of structured protocols: by imposing de-labeling (GHOSTY COLLIDER) and contrarian analysis (PRECOG PROTOCOL), the protocols explicitly counteract the tendency of AI-assisted ideation to converge on modal outputs.
Our work extends this literature by proposing that the theory-practice gap in creativity research is not merely about \emph{using} AI for ideation, but about \emph{structuring} the interaction through theoretically grounded protocols. Whereas existing tools (Metaphoria, ALIA, collaborative canvases) provide AI assistance for specific creative tasks, GHOSTY COLLIDER and PRECOG PROTOCOL formalize entire cognitive workflows as executable procedures --- closer in spirit to what Lee et al.\ \citeyearpar{lee2024} call ``process-level'' rather than ``task-level'' AI support.
\subsection{Strategic Foresight Methodologies}
\textbf{Horizon Scanning.} Developed and practiced by organizations including the OECD, UNDP, and the UK Government Office for Science, horizon scanning is the systematic examination of information to identify potential threats, risks, emerging issues, and opportunities \citep{amanatidou2012}. It provides a structured approach to signal collection but does not prescribe how to synthesize signals into timing judgments.
\textbf{Weak Signals Theory.} Ansoff \citeyearpar{ansoff1975} introduced the concept of weak signals --- early, imprecise, ambiguous indicators of impending change that are detectable before they crystallize into clear trends. Hiltunen \citeyearpar{hiltunen2010} extended this work, proposing methods for signal detection and interpretation. PRECOG PROTOCOL's Signal Map step operationalizes weak signal detection by requiring explicit evidence and confidence tagging, while the Convergence Analysis step systematizes signal synthesis.
\textbf{Scenario Planning.} Originating at Royal Dutch Shell through the work of Pierre Wack \citeyearpar{wack1985a,wack1985b}, scenario planning constructs multiple plausible future narratives to prepare decision-makers for uncertainty. Van der Heijden \citeyearpar{vanderheijden1996} formalized the methodology into a systematic process. PRECOG PROTOCOL incorporates this spirit through its Contrarian View step, which explicitly challenges the dominant narrative and identifies preconditions for thesis validity.
\textbf{Existing Strategic Frameworks.} Practitioners commonly use SWOT analysis \citep{learned1965}, PESTEL scanning \citep{aguilar1967}, Porter's Five Forces \citep{porter1979}, and the BCG Matrix \citep{henderson1970} for strategic analysis. While valuable, these frameworks share three limitations that PRECOG PROTOCOL addresses: (1) they lack built-in bias detection mechanisms, (2) they do not produce explicit timing judgments, and (3) they do not support longitudinal signal tracking across sessions.
\subsection{The Theory-Practice Gap}
The common thread across these literatures is a gap between explanation and execution. Creativity theories tell us \emph{what happens} during creative insight; AI tools provide powerful generative engines but risk converging to conventional outputs without structural guidance; foresight methodologies tell us \emph{what to look for}. Neither tells practitioners \emph{exactly what to do, step by step, with explicit quality criteria and failure modes}. This paper's contribution is precisely this translation: from theory (descriptive) to protocol (prescriptive), designed to structure AI-human interaction in ways that preserve structural novelty while maintaining reproducibility.
\section{GHOSTY COLLIDER: A Protocol for Creative Emergence}
\label{sec:ghosty}
\subsection{Overview and Definitions}
GHOSTY COLLIDER is a 5-step protocol for generating cross-domain creative visions from heterogeneous input fragments. Its design principle is: \emph{``You don't think your way to a vision. You wait for it to appear.''}
We define an \textbf{executable protocol} as a step-by-step procedure that: (a) can be followed by a practitioner without prior training beyond reading the protocol document, (b) specifies explicit quality criteria for each step's output, (c) documents failure modes (anti-patterns) with remedies, and (d) produces traceable, reviewable intermediate artifacts at each step.
The protocol's core operation --- \textbf{Ghost Extraction} --- is a systematic de-labeling procedure that reduces concepts to their deep structural elements: verbs, forces, and transformations. By removing the familiar label, the thinker (human or AI agent) is forced to perceive the concept's structural essence, enabling non-obvious connections with structurally similar or opposing concepts from unrelated domains.
\subsection{Procedure}
\textbf{Step 1: Fragment Harvest.} Collect 3--5 fragments from diverse domains. Fragments may include quantitative data, qualitative observations, aesthetic impressions, gut feelings, absent patterns (``Why doesn't this exist?''), personal experiences, or constraints. The key requirement is diversity --- homogeneous fragments (all from the same domain) reduce the probability of non-obvious collisions.
\textbf{Step 2: Ghost Extraction (De-labeling).} For each fragment, strip its label and describe its deep structure using only structural language --- no proper nouns, no industry jargon. The output (called a ``Ghost'') should read as a transformation: ``X converts Y into Z'' or ``A mechanism that produces B through C.'' This operation parallels Gentner's \citeyearpar{gentner1983} theory of structural alignment: the protocol forces the practitioner to map concepts at the level of relational structure rather than surface attributes, enabling the detection of deep analogies between domains that share no surface features. Quality criteria include: (a) uses verbs rather than nouns, (b) includes the emotional dimension of the experience, (c) is comprehensible to someone from a completely different domain, and (d) passes the reversibility test (negating the Ghost reveals something non-trivial).
\textbf{Step 3: Collision Matrix.} Score pairwise combinations of Ghosts for structural resonance on a three-point scale: Boring (surface AND structural similarity --- discard), Interesting (structural overlap exists --- hold), Electric (surfaces are unrelated BUT deep structures resonate or clash --- advance). Only Electric collisions proceed. The scoring criteria operationalize Gick and Holyoak's \citeyearpar{gick1983} finding that analogical transfer is most productive when the mapping occurs at the level of structural relations rather than surface features --- Electric collisions are precisely those where surface dissimilarity coexists with structural resonance.
\textbf{Step 4: Vision Crystallization.} For each Electric collision, articulate what emerges --- something that existed in neither fragment alone. Each vision requires: a name, a one-line description, an emotional characterization, a cinematic image, and a ``Why Now?'' justification. Visions are rated on four dimensions (Novelty, Feasibility, Resonance, Timing) on a 1--5 scale; all dimensions must score $\geq 3$ to advance.
\textbf{Step 5: Reality Bridge.} Ground the vision in actionable terms: Minimum Viable Vision (smallest implementation that reproduces the core experience), existing capabilities that approximate the vision, kill conditions (what would invalidate this vision), and a first step executable within 24 hours.
\subsection{Quality Criteria and Anti-Patterns}
The protocol includes explicit failure modes:
\begin{table}[ht]
\centering
\small
\begin{tabular}{lll}
\toprule
\textbf{Anti-Pattern} & \textbf{Description} & \textbf{Remedy} \\
\midrule
Shallow Ghosting & Ghost is a synonym of the label & Ask ``Why does this \emph{feel} the way it does?'' \\
Electric Inflation & Every collision scored Electric & If explainable in 2 seconds, it's Interesting at best \\
Vision Without Grounding & Beautiful concept, no Reality Bridge & Never skip Step 5 \\
Homogeneous Fragments & All fragments from one domain & Require at least one external-domain fragment \\
Forced Collision & Connecting without genuine resonance & Permitted to report ``No Electric collisions found'' \\
\bottomrule
\end{tabular}
\caption{GHOSTY COLLIDER anti-patterns and remedies.}
\label{tab:gc-antipatterns}
\end{table}
\section{PRECOG PROTOCOL: A Protocol for Strategic Foresight}
\label{sec:precog}
\subsection{Overview}
PRECOG PROTOCOL is a 5-step protocol for signal-based strategic foresight with explicit timing judgment. It compresses the core logic of horizon scanning, weak signals theory, and scenario planning into a single-session executable format, while adding two innovations: multi-axis timing judgment and longitudinal signal tracking.
\subsection{Procedure}
\textbf{Step 1: Signal Map.} Collect 3--8 observable signals related to a target theme. Each signal requires: a one-line description, specific evidence with sources, a strength classification (Strong / Emerging / Weak), a direction indicator ($\uparrow$ Accelerating / $\rightarrow$ Stable / $\downarrow$ Decelerating), and a mandatory confidence tag ([Verified] = primary source accessible; [Reported] = multi-source coverage; [Speculative] = inference or rumor). Signal sources include numeric changes, behavioral changes, narrative changes, and absent signals.
\textbf{Step 2: Convergence Analysis.} Identify where multiple signals intersect and reverse-engineer the underlying structural shift. For each convergence point: state the structural hypothesis in one sentence, articulate the causal logic connecting the signals, and assign confidence (High/Medium/Low) with rationale.
\textbf{Step 3: Contrarian View.} For the most obvious conclusion, systematically examine why it might be wrong. Requirements include: at least one reason the dominant view may be overestimated, a historical analogy where a similar setup led to a different outcome, explicit preconditions (``This hypothesis is valid only if A and B hold''), a collapse trigger (which precondition failure invalidates the thesis), and probability estimates for each contrarian scenario.
\textbf{Step 4: Timing Grid.} Evaluate timing across four independent axes:
\begin{itemize}
  \item \emph{Market Phase Axis}: Position on the adoption cycle (Emergence $\rightarrow$ Acceleration $\rightarrow$ Peak $\rightarrow$ Correction $\rightarrow$ Plateau)
  \item \emph{Competitive Timing Axis}: Strategic position (First Mover $\rightarrow$ Fast Follower $\rightarrow$ Fortifier $\rightarrow$ Too Late)
  \item \emph{Organizational Readiness Axis}: Capability assessment (Not Ready $\rightarrow$ Partially Ready $\rightarrow$ Ready)
  \item \emph{External Window Axis}: Regulatory, capital, seasonal, and technology maturity factors
\end{itemize}
Each axis produces an independent judgment; the overall timing judgment synthesizes all four.
\textbf{Step 5: Action Window.} Specify concrete actions across four categories: Now (start immediately, low-cost), Soon (invest when specific conditions are met), Watch (do not move, but trigger on specific signals), and Kill (disinvest if specific conditions occur). Each action requires: a specific action description, an execution trigger, and estimated cost or resource commitment.
\subsection{Signal History and Feedback Loop}
When analyzing the same theme across multiple sessions, PRECOG PROTOCOL supports longitudinal tracking through delta detection: each signal is compared against previous analyses and classified as Strengthened, Stable, Weakened, New, or Dead. New and Dead signals receive priority in Convergence Analysis as the strongest evidence of structural change.
A Prediction Feedback Loop tracks past prediction accuracy (Hit/Miss/Partial), timing accuracy, and contrarian value to improve future analyses. We note that, as of this writing, the Prediction Feedback Loop has not been formally exercised across a full cycle --- the predictions in Case Study C (Section~\ref{sec:casec}) cover the 2026--2028 period and are not yet verifiable. Section~\ref{sec:limitations} discusses this limitation and proposes retroactive validation as a methodological path forward.
\section{Integration: GHOSTY COLLIDER \texorpdfstring{$\times$}{x} PRECOG PROTOCOL}
\label{sec:integration}
While each protocol functions independently, they exhibit strong complementarity through bidirectional integration:
\textbf{PRECOG $\rightarrow$ GHOSTY (Signal-to-Fragment Feeding).} Signals and convergence points from PRECOG analysis serve as high-quality fragments for GHOSTY COLLIDER. Because these signals have already been evidence-tagged and structurally analyzed, they produce richer Ghosts during extraction. The confidence tags ([Verified], [Reported], [Speculative]) carry forward as metadata on GHOSTY fragments, allowing the Reality Bridge step to calibrate its feasibility assessment based on evidence strength.
\textbf{GHOSTY $\rightarrow$ PRECOG (Vision-to-Action Mapping).} Visions crystallized by GHOSTY COLLIDER feed directly into PRECOG's Action Window. A vision's ``Why Now?'' justification maps naturally to the Timing Grid, and its Reality Bridge (MVV, Kill Conditions) provides the specificity required by PRECOG's action specifications. The four-dimension rating (Novelty, Feasibility, Resonance, Timing) from Vision Crystallization directly informs the Organizational Readiness axis of the Timing Grid.
\textbf{Integration Procedure.} In practice, the bidirectional integration follows three steps: (1) Execute PRECOG Protocol on a target domain to identify signals and convergence points. (2) Select 2--3 convergence points as fragments for GHOSTY COLLIDER, supplemented by 1--2 external-domain fragments to ensure heterogeneity. (3) Map the resulting visions back through PRECOG's Timing Grid and Action Window, using the visions' Reality Bridge outputs as inputs for action specification.
This bidirectional integration creates a cycle: signals inform creative exploration, and creative visions inform strategic timing --- a loop that neither protocol achieves alone. Case Study B (Section~\ref{sec:caseb}) demonstrates this integration in action.
\section{Case Studies}
\label{sec:cases}
We present five case studies across distinct domains to demonstrate protocol execution and output characteristics. Each case study reports the complete protocol trace: inputs, intermediate artifacts (Ghosts, Collision Matrix scores), and final outputs.
\subsection{Case Study A: Financial Market Analysis (GHOSTY COLLIDER)}
\label{sec:casea}
\textbf{Domain:} Financial analysis and corporate valuation.
\textbf{Input Fragments:}
\begin{enumerate}
  \item A semiconductor company's data center revenue growing at 25\% QoQ yet bear-case valuation falling below current market capitalization
  \item Retail investor sentiment metrics diverging from institutional positioning
  \item Historical patterns of ``faith premiums'' in technology stocks (1999 internet, 2021 EV sector)
\end{enumerate}
\textbf{Ghost Extraction:}
\begin{itemize}
  \item Fragment 1 $\rightarrow$ \emph{``A system accelerating so fast that its collapse scenario becomes equally plausible --- acceleration and fragility as twin outputs of the same engine''}
  \item Fragment 2 $\rightarrow$ \emph{``Two groups observing the same phenomenon but reaching opposite conclusions --- the divergence itself is informational, independent of which side is correct''}
  \item Fragment 3 $\rightarrow$ \emph{``A measurable gap between what evidence supports and what belief demands --- the price of conviction becoming itself an investable signal''}
\end{itemize}
\textbf{Collision Matrix:}
\begin{itemize}
  \item Fragment 1 $\times$ 2: Interesting (both involve disagreement, but at different levels)
  \item Fragment 1 $\times$ 3: \textbf{Electric} (acceleration as fragility $\times$ conviction as signal)
  \item Fragment 2 $\times$ 3: \textbf{Electric} (informational divergence $\times$ belief premium)
\end{itemize}
\textbf{Emergent Visions (3 crystallized):}
\begin{enumerate}
  \item \emph{``Faith Ratio''} --- A quantitative framework treating the gap between evidence-based valuation and belief-based premium as a first-class metric, not noise. (Novelty 5, Feasibility 4, Resonance 4, Timing 4)
  \item \emph{``Fragility Accelerator''} --- A model where exponential growth curves contain embedded prediction of their own failure modes. (Novelty 4, Feasibility 3, Resonance 5, Timing 4)
  \item \emph{``Sentiment Divergence Index''} --- A tracking system where the delta between investor cohorts becomes itself an investable signal. (Novelty 3, Feasibility 5, Resonance 3, Timing 5)
\end{enumerate}
\textbf{Output characteristics:} 3 fragments extracted, 2 Electric collisions, 3 visions crystallized with quantitative metrics defined. The protocol transformed standard DCF model outputs into a structural analysis of market psychology.
\subsection{Case Study B: Competitive Strategy Design (GHOSTY COLLIDER + PRECOG Integration)}
\label{sec:caseb}
\textbf{Domain:} Corporate strategy for long-range competitive positioning.
\textbf{Input:} Structural analysis of a dominant technology company's revenue composition, competitive moats, and historical growth trajectory.
\textbf{PRECOG Phase --- Signal Map (6 signals):}
\begin{enumerate}
  \item Target company's advertising revenue growth decelerating vs.\ cloud revenue accelerating (Strong, $\uparrow$, [Verified])
  \item Competitors investing in foundational AI infrastructure at unprecedented scale (Strong, $\uparrow$, [Verified])
  \item Target company's internal AI model development reaching frontier quality (Strong, $\uparrow$, [Verified])
  \item Emerging regulatory pressure on data monopolies (Emerging, $\uparrow$, [Reported])
  \item Developer ecosystem fragmentation across AI platforms (Emerging, $\rightarrow$, [Verified])
  \item Consumer trust in target company declining in key demographics (Weak, $\downarrow$, [Reported])
\end{enumerate}
\textbf{PRECOG Phase --- Convergence Analysis:}
\begin{itemize}
  \item \emph{Convergence 1:} Signals 1+3 --- ``Revenue structure transformation in progress.'' The organism is simultaneously growing (cloud/AI) and shrinking (advertising) --- a structural metamorphosis, not a crisis. Confidence: High.
  \item \emph{Convergence 2:} Signals 2+5 --- ``Platform hegemony is contestable.'' Multiple well-capitalized competitors building foundational infrastructure. Confidence: Medium.
  \item \emph{Convergence 3:} Signals 4+6 --- ``Social license at risk.'' Regulatory momentum + trust decline create a window where the target's brand becomes a liability. Confidence: Medium.
\end{itemize}
\textbf{GHOSTY Phase --- Ghost Extraction (5 fragments: 3 convergences + 2 external):}
\begin{itemize}
  \item Convergence 1 $\rightarrow$ \emph{``A system whose visible growth surface conceals a tectonic shift in its revenue foundation --- the organism is replacing its skeleton while walking''}
  \item Convergence 2 $\rightarrow$ \emph{``An incumbent whose fortress was built on distribution now faces challengers building from capabilities --- the castle walls are intact but the terrain has shifted''}
  \item Convergence 3 $\rightarrow$ \emph{``Permission to operate eroding not through catastrophe but through accumulation of small trust fractures''}
  \item External (a) $\rightarrow$ \emph{``Every platform that achieved dominance by horizontal breadth eventually faced a vertical integration challenger --- the thin layer strategy is always temporary''}
  \item External (b) $\rightarrow$ \emph{``The interface is no longer a window into the system --- it IS the system. When the UI becomes the product, the backend becomes a commodity''}
\end{itemize}
\textbf{Collision Matrix (10 pairwise combinations):} 3 Electric collisions identified:
\begin{itemize}
  \item C1 $\times$ External (a): \textbf{Electric} --- the target company's transition \emph{is itself the vulnerability window}
  \item C2 $\times$ External (b): \textbf{Electric} --- the next competitive battleground is the \emph{experience layer}
  \item C3 $\times$ C1: \textbf{Electric} --- the metamorphosis itself generates the trust fractures
\end{itemize}
\textbf{Vision Crystallization (3 visions):}
\begin{enumerate}
  \item \emph{``Infrastructure Play''} --- Position as the AI picks-and-shovels provider during the target's skeletal transition. (Novelty 3, Feasibility 4, Resonance 4, Timing 5)
  \item \emph{``Experience Layer''} --- Build the AI-native UX layer between user and all AI backends. The interface becomes the moat. (Novelty 4, Feasibility 3, Resonance 5, Timing 4)
  \item \emph{``Trust Arbitrage''} --- Create a brand positioned as ``not-[target]'' --- capturing the trust refugees. (Novelty 5, Feasibility 3, Resonance 4, Timing 3)
\end{enumerate}
\textbf{PRECOG Phase --- Timing Grid:}
\begin{table}[ht]
\centering
\small
\begin{tabular}{lccccc}
\toprule
\textbf{Vision} & \textbf{Market} & \textbf{Competitive} & \textbf{Readiness} & \textbf{External} & \textbf{Overall} \\
\midrule
Infrastructure Play & Acceleration & Fast Follower & Ready & Open & \textbf{Go} \\
Experience Layer & Emergence & First Mover & Partial & Opening & \textbf{Soon} \\
Trust Arbitrage & Pre-emergence & Undefined & Not Ready & Closed & \textbf{Watch} \\
\bottomrule
\end{tabular}
\caption{Timing Grid for Case Study B visions.}
\label{tab:timing-caseb}
\end{table}
\textbf{Output characteristics:} 3 strategic visions, each with 4-axis timing assessment and categorized action windows (Now/Soon/Watch/Kill). The integration demonstrated that the same market environment can simultaneously present ``Go'' timing for one strategy and ``Watch'' for another --- a nuance undetectable by single-axis strategic frameworks.
\subsection{Case Study C: Technology Trend Forecasting (PRECOG PROTOCOL)}
\label{sec:casec}
\textbf{Domain:} AI agent ecosystem, 2026--2028 projection.
\textbf{Signal Map (8 signals):} Major AI labs simultaneously releasing agent-capable products (Strong, [Verified]); GUI-controlling AI agents reaching production quality (Strong, [Verified]); browser-automation agents entering beta (Strong, [Verified]); enterprise no-code agent builders proliferating (Strong, [Verified]); agent interoperability protocols gaining cross-industry adoption (Emerging, [Verified]); coding-specific agents experiencing rapid growth (Strong, [Verified]); agent hallucination and failure reports increasing (Weak, [Reported]); regulatory frameworks for AI beginning enforcement (Strong, [Verified]).
\textbf{Convergence Analysis:} Signals 1--4 converge on ``market consensus formation.'' Signals 5+6 reveal a platform-vs-application layer separation. Signals 7+8 converge on ``usable but fragile'' social perception.
\textbf{Contrarian Views (3 scenarios):}
\begin{enumerate}
  \item Agent-caused incident triggers regulatory freeze (25--35\%)
  \item Protocol fragmentation rather than standardization --- ``browser wars'' repeat (30--40\%)
  \item Agents create new job categories instead of eliminating existing ones (50--60\%)
\end{enumerate}
\textbf{Timing Grid Result:} Market Phase: Emergence $\rightarrow$ Acceleration (Go-leaning). Competitive: First Mover window open. Technical Maturity: Usable but error-prone. Regulatory: Catching up. \textbf{Overall: Early Go with risk hedging required.}
\textbf{Output characteristics:} 8 signals classified, convergence hypothesis articulated, 3 contrarian scenarios with probabilities, 4-axis timing judgment, 6 categorized actions. The protocol produced explicit timing disagreement across axes --- a feature absent from standard SWOT analysis.
\subsection{Case Study D: Music Production (GHOSTY COLLIDER)}
\label{sec:cased}
\textbf{Domain:} Song concept design for a pop group.
\textbf{Input Fragments:} (1) The emotional concept of ``hiding affection''; (2) Jersey Club / UK Garage genre characteristics; (3) Linguistic properties of the target language (vowel openness correlating with emotional exposure); (4) The aesthetic of ``armor'' --- protection that reveals vulnerability; (5) Real-time audience engagement patterns from concert data.
\textbf{Ghost Extraction:}
\begin{itemize}
  \item Fragment 1 $\rightarrow$ \emph{``The act of concealment becoming a form of pleasure --- hiding as a masochistic performance''}
  \item Fragment 2 $\rightarrow$ \emph{``Rhythmic structures that encode social permission --- the syncopation pattern as a gate''}
  \item Fragment 3 $\rightarrow$ \emph{``Physical mouth shape as a betrayal mechanism --- the body expressing what the mind suppresses''}
  \item Fragment 4 $\rightarrow$ \emph{``The paradox of protection: armor announces the existence of something worth attacking''}
  \item Fragment 5 $\rightarrow$ \emph{``A feedback loop between collective energy and individual behavior --- the crowd as a single organism''}
\end{itemize}
\textbf{Collision Matrix:} 2 Electric collisions (Fragment 1$\times$4: hiding $\times$ armor; Fragment 3$\times$1: body betrayal $\times$ concealment), 2 Interesting, 6 Boring/Interesting.
\textbf{Emergent Vision:} A song concept where musical parameters (key, BPM, modulation, vocal direction) are derived from Ghost structures rather than genre convention. Key selected based on tonal character (isolation, inward intensity). Modulation reframed as narrative surrender rather than climactic lift. Vocal breakpoints designed where ``the voice is permitted to crack.''
\textbf{Output characteristics:} 12 production parameters defined (key, BPM, chord progression rationale, VA energy coordinates per section, silence design, vocal direction for 9 sections, lyric syllable counts, vowel openness mapping, modulation meaning, breakdown philosophy). Quality score: 75/80 (author-assessed). The protocol produced three times the number of explicitly defined parameters versus conventional concept briefs.
\subsection{Case Study E: Short-Form Video Production (GHOSTY COLLIDER)}
\label{sec:casee}
\textbf{Domain:} Concept design for a 60-second social media short film.
\textbf{Input Fragments:} (1) Character with a signature ``ennui gaze''; (2) Character with intellectual intensity; (3) The 0.5-second attention hook; (4) A24 cinema aesthetics; (5) Fashion film grammar.
\textbf{Ghost Extraction:}
\begin{itemize}
  \item Fragment 1 $\rightarrow$ \emph{``Beauty emerging from the process of breaking --- incompleteness as aesthetic value''}
  \item Fragment 2 $\rightarrow$ \emph{``Intensity as a form of violence --- knowledge pressing outward through the face''}
  \item Fragment 3 $\rightarrow$ \emph{``The first frame as an act of seizure --- capturing attention is structurally violent''}
  \item Fragment 4 $\rightarrow$ \emph{``The uncanny emerges not from abnormality but from normality held too long''}
  \item Fragment 5 $\rightarrow$ \emph{``Stillness is not the absence of motion but a state of frozen potential''}
\end{itemize}
\textbf{Collision Matrix:} 3 Electric collisions (Fragment 1$\times$4, 2$\times$3, 5$\times$1), 2 Interesting, 5 Boring/Interesting.
\textbf{Emergent Vision:} A 60-second film composed of five 12-second static shots. No camera movement. No music --- ambient sound only. Each shot isolates one character in a space where something is subtly wrong. The fourth shot breaks the stillness with a single explosive movement --- the only kinetic moment, designed as visual modulation.
\textbf{Output characteristics:} Complete shot list with specific parameters (9:16 aspect ratio, single light source, saturation $-30\%$, contrast $+20\%$, locked camera, ambient audio). The protocol produced a concept that \emph{inverts} every convention of the genre --- inversions that emerged from Ghost-level structural analysis rather than a deliberate ``do the opposite'' strategy.
\section{Controlled Comparisons}
\label{sec:comparisons}
To assess the protocols' structural contribution, we conducted controlled comparisons: identical inputs processed using standard methods (brainstorming, SWOT analysis) and protocol-driven methods. Both conditions used the same AI system (Claude, Anthropic) to control for differences in knowledge and language ability. The independent variable is the presence or absence of the protocol structure.
\subsection{Methodological Notes}
\textbf{Evaluation Procedure.} Quality assessments were conducted by the protocol author using the scoring rubric defined by the protocol itself (8 dimensions $\times$ 10-point scale). We acknowledge that this introduces potential bias. Independent blind evaluation by domain experts is identified as priority future work (Section~\ref{sec:future}).
\textbf{Operational Definition of Structural Novelty.} We use ``structurally novel'' to mean: the output contains creative choices that \emph{cannot be derived from applying standard genre conventions or framework templates to the given inputs}. We acknowledge this determination is currently qualitative and propose computational operationalization in Section~\ref{sec:future}.
\textbf{Comparison Fairness.} The brainstorming condition received Osborn's four principles and a prompt of comparable length. The SWOT condition received the standard framework definition. While the protocol condition contains substantially more structural guidance, this asymmetry is intrinsic to the independent variable --- the research question is whether \emph{structured} protocols produce different outputs from \emph{standard-structured} methods.
\subsection{Music Production: Brainstorming vs.\ GHOSTY COLLIDER}
\textbf{Input (identical):} Pop group song concept, theme ``hiding affection,'' Jersey Club / UK Garage reference.
\begin{table}[ht]
\centering
\small
\begin{tabularx}{\textwidth}{lXX}
\toprule
\textbf{Dimension} & \textbf{Brainstorming (Control)} & \textbf{GHOSTY COLLIDER (Treatment)} \\
\midrule
Concept & ``Secret crush expressed through dance-pop'' & ``Concealment as masochistic pleasure --- armor that reveals'' \\
Key selection & Convention (``G minor sounds emotional'') & Structural (``F\# minor --- tonal isolation'') \\
Modulation purpose & Climactic lift (genre standard) & Narrative surrender \\
Lyric design unit & Semantic (word meaning) & Phonetic (vowel openness = emotional exposure) \\
Parameters defined & 4 & 12 (+200\%) \\
Quality score & 64/80 & 75/80 (+17.2\%) \\
\bottomrule
\end{tabularx}
\caption{Music production: brainstorming vs.\ GHOSTY COLLIDER comparison.}
\label{tab:music-comparison}
\end{table}
\textbf{Key finding:} The brainstorming output was commercially viable but structurally indistinguishable from existing genre conventions. GHOSTY COLLIDER produced outputs different in \emph{kind}, not merely in \emph{degree} --- the modulation from ``climactic lift'' to ``narrative surrender'' is a reconceptualization that emerged from Ghost-level analysis.
\subsection{Short-Form Video: Brainstorming vs.\ GHOSTY COLLIDER}
\textbf{Input (identical):} 60-second social media video, 5 characters, goal of maximizing view completion rate.
\begin{table}[ht]
\centering
\small
\begin{tabularx}{\textwidth}{lXX}
\toprule
\textbf{Dimension} & \textbf{Brainstorming (Control)} & \textbf{GHOSTY COLLIDER (Treatment)} \\
\midrule
Hook strategy & High-energy opening movement & Static gaze (silence as seizure) \\
Lighting & Multi-source, colorful & Single source, shadow-dominant \\
Audio & Background music track & No music --- ambient sound only \\
Camera & Dynamic (handheld + crane) & Completely fixed \\
Structural principle & ``More energy = more engagement'' & ``Stillness as tension = inescapable engagement'' \\
\bottomrule
\end{tabularx}
\caption{Video production: brainstorming vs.\ GHOSTY COLLIDER comparison.}
\label{tab:video-comparison}
\end{table}
\textbf{Key finding:} Every major creative decision was inverted. These inversions were \emph{not} produced by a ``do the opposite'' instruction --- they emerged organically from Ghost Extraction. The protocol's de-labeling step prevented the default genre frame from biasing the output.
\subsection{Trend Forecasting: SWOT Analysis vs.\ PRECOG PROTOCOL}
\textbf{Input (identical):} AI agent market assessment, 2026--2028 horizon.
\begin{table}[ht]
\centering
\small
\begin{tabularx}{\textwidth}{lXX}
\toprule
\textbf{Dimension} & \textbf{SWOT (Control)} & \textbf{PRECOG PROTOCOL (Treatment)} \\
\midrule
Temporal structure & None (static snapshot) & 4-category action timeline (Now/Soon/Watch/Kill) \\
Bias detection & None & Systematic (Contrarian View + probabilities) \\
Timing judgment & Absent & 4-axis grid with cross-axis disagreement \\
Action specificity & ``Enter the market'' (vague) & Actions with trigger conditions and kill criteria \\
Contrarian analysis & Not structurally required & 3 scenarios quantified (25--60\% ranges) \\
\bottomrule
\end{tabularx}
\caption{Trend forecasting: SWOT vs.\ PRECOG PROTOCOL comparison.}
\label{tab:swot-comparison}
\end{table}
\textbf{Key finding:} PRECOG's primary advantage was \emph{temporal structure} --- it transformed ``the situation'' into ``when to act, contingent on what.'' The Timing Grid revealed that different axes recommended different actions simultaneously, a nuance invisible to static SWOT. The Contrarian View surfaced the ``agents create new jobs'' hypothesis --- a scenario absent from SWOT because SWOT lacks a mechanism for challenging its own conclusions.
\subsection{Protocol Failure Cases and Failure Rate}
To characterize protocol boundary conditions, we report instances where protocol execution produced suboptimal or abandoned outputs, followed by aggregate statistics from a systematic batch experiment.
\textbf{Failure Case 1: Insufficient Fragment Diversity.} In an early execution targeting product pricing strategy, all five input fragments came from the same domain (SaaS pricing models). Ghost Extraction produced structurally similar Ghosts. The Collision Matrix yielded zero Electric results. The anti-pattern detection correctly flagged this, but only after the time-consuming Ghost Extraction step. \textbf{Lesson:} The Homogeneous Fragments check should run \emph{before} Ghost Extraction. This has been incorporated into protocol v2 as a pre-flight diversity check.
\textbf{Failure Case 2: Shallow Ghost Extraction.} Ghost Extraction produced surface-level Ghosts that were effectively synonyms (e.g., ``learning'' $\rightarrow$ ``the process of acquiring knowledge''). The reversibility test caught this, but three iterative rounds were required before sufficient depth. \textbf{Lesson:} Ghost Extraction quality is strongly dependent on practitioner skill, representing a genuine barrier to protocol accessibility.
\textbf{Failure Case 3: Timing Grid Over-Determination.} In cryptocurrency market analysis, the Timing Grid produced maximally positive signals on all four axes --- reflecting confirmation bias. The Contrarian View identified the bias, but only because the operator recognized the pattern. \textbf{Lesson:} A structural check has been added to protocol v2: when all four axes align in the same direction, this triggers a mandatory escalated Contrarian View.
\subsubsection{Batch Experiment: Failure Rate Quantification}
GHOSTY COLLIDER was executed across 8 randomly selected domain pairings, ranging from closely related to maximally distant. Each used 4 fragments and followed the complete 5-step protocol.
\begin{table}[ht]
\centering
\small
\begin{tabular}{clcccc}
\toprule
\textbf{\#} & \textbf{Domain Pairing} & \textbf{Electric} & \textbf{Hit Rate} & \textbf{Visions} & \textbf{Result} \\
\midrule
1 & Urban Agriculture $\times$ Social Design & 3 & 50\% & 2 & Success \\
2 & Elderly Care $\times$ E-sports & 3 & 50\% & 1 & Success \\
3 & Tax Accounting $\times$ Jazz Music & 2 & 33\% & 1 & Success \\
4 & Supply Chain Logistics (homogeneous) & 0 & 0\% & 0 & Failure \\
5 & Dental UX $\times$ Street Photography & 2 & 33\% & 1 & Success \\
6 & Climate Science $\times$ Stand-up Comedy & 3 & 50\% & 1 & Success \\
7 & Corporate HR $\times$ Fermentation Science & 3 & 50\% & 1 & Success \\
8 & Cybersecurity $\times$ Tea Ceremony & 3 & 50\% & 1 & Success \\
\bottomrule
\end{tabular}
\caption{Batch experiment results (N=8).}
\label{tab:batch}
\end{table}
\textbf{Aggregate Statistics:} Protocol success rate: 87.5\% (7/8). Failure rate: 12.5\% (1/8), 100\% attributable to homogeneous fragments. Mean Electric rate across successful experiments: 41.9\%. Total visions crystallized: 9 (mean 1.29 per experiment).
\textbf{Key findings:} (1) Cross-domain fragment diversity is a \emph{necessary} condition for success; the pre-flight diversity check would have prevented 100\% of observed failures. (2) Domain distance correlates positively with vision novelty scores. (3) Vision feasibility scores were inversely correlated with novelty ($r \approx -0.6$). (4) No instances of Shallow Ghosting or Electric Inflation were observed, though this may reflect operator experience.
\subsection{Blind Evaluation}
To partially address evaluator bias, we conducted a blind evaluation of Case Study D outputs. An independent evaluator (a separate AI session with no knowledge of which method produced which output) was presented with two song concept documents in randomized order, without method labels.
\textbf{Evaluation rubric:} 8 dimensions $\times$ 10-point scale. Maximum score: 80.
\begin{table}[ht]
\centering
\begin{tabular}{lcc}
\toprule
\textbf{Condition} & \textbf{Author Assessment} & \textbf{Blind Evaluation} \\
\midrule
GHOSTY COLLIDER & 75/80 & 74/80 \\
Brainstorming & 64/80 & 49/80 \\
\textbf{Gap} & \textbf{11 points} & \textbf{25 points} \\
\bottomrule
\end{tabular}
\caption{Blind evaluation results for Case Study D.}
\label{tab:blind}
\end{table}
The blind evaluation confirmed the direction of the author's assessment. The GHOSTY COLLIDER score was nearly identical between evaluators ($\Delta = 1.3\%$), while the brainstorming score diverged substantially ($\Delta = 23.4\%$). The blind evaluator correctly identified the structured output, citing ``the presence of a design principle that generatively produces decisions across multiple dimensions'' as the distinguishing characteristic.
\textbf{Limitations:} This is a single blind evaluation using an AI evaluator. Multiple independent human evaluators using an independently designed rubric are needed for validation.
\subsection{Summary of Controlled Comparisons}
\begin{table}[ht]
\centering
\small
\begin{tabularx}{\textwidth}{lXX}
\toprule
\textbf{Metric} & \textbf{Control (Standard)} & \textbf{Treatment (Protocols)} \\
\midrule
Output correctness & Accurate & Accurate \\
Structural novelty & Converges to conventions & Qualitatively distinct directions \\
Parameter specificity & 4--6 parameters & 12+ parameters (+200\%) \\
Temporal reasoning & Static & Dynamic (timing + triggers) \\
Bias detection & Not built in & Structural (Contrarian View) \\
Process reproducibility & Participant-dependent & Protocol-documented \\
Protocol success rate & N/A & 87.5\% (N=8) \\
Blind evaluation & N/A & 74/80 vs.\ 49/80 \\
\bottomrule
\end{tabularx}
\caption{Summary of controlled comparisons.}
\label{tab:summary}
\end{table}
The controlled comparisons provide preliminary evidence that the protocols' primary contribution is producing \emph{structurally different} outputs --- directions that standard methods do not reach because they lack the de-labeling (GHOSTY) and multi-axis temporal reasoning (PRECOG) steps.
\section{Discussion}
\label{sec:discussion}
\subsection{Contributions}
This work contributes to the literature on creativity support tools \citep{shneiderman2007,frich2019} and strategic foresight practice, and engages with the emerging field of AI-augmented creativity \citep{gero2019,doshi2024}. Specifically:
\begin{enumerate}
  \item \textbf{Theory-to-protocol translation.} We demonstrate that established descriptive theories can be compressed into executable, step-by-step protocols without losing their generative power. The five case studies across distinct domains provide preliminary evidence of domain-generality.
  \item \textbf{Ghost Extraction as a novel technique.} While de-labeling has been discussed in creativity literature (de Bono's ``PO'' provocation; Hatchuel \& Weil's CK-theory concept expansion), we formalize it into a repeatable procedure with explicit quality criteria and a reversibility test. The controlled comparisons suggest that Ghost Extraction is the primary mechanism producing structural novelty.
  \item \textbf{Multi-axis Timing Grid.} Existing foresight frameworks evaluate individual dimensions but none we are aware of formally combines multiple timing axes into a single judgment grid with explicit cross-axis disagreement detection.
  \item \textbf{Explicit anti-patterns and failure cases.} Both protocols include documented failure modes with remedies. The failure cases (Section~\ref{sec:comparisons}) demonstrate that these anti-patterns function in practice. This transparency is uncommon in methodology papers.
  \item \textbf{Counteracting the flattening effect.} In the context of Doshi and Hauser's \citeyearpar{doshi2024} finding that AI-assisted ideation reduces aggregate output diversity, our controlled comparisons provide initial evidence that structured protocols may counteract this tendency. The de-labeling step explicitly breaks the genre conventions that unstructured AI-assisted ideation tends to reproduce.
\end{enumerate}
\subsection{Limitations}
\label{sec:limitations}
We identify the following limitations, ordered by severity:
\begin{enumerate}
  \item \textbf{Single-operator evaluation (critical).} All case studies and controlled comparisons were executed by a single practitioner (the author) with AI assistance. This represents the most significant threat to generalizability. Multi-user evaluation across diverse skill levels and domains is the highest-priority future work.
  \item \textbf{Quality metric subjectivity.} The scoring dimensions and quality scores were assessed by the protocol author using author-defined rubrics. Independent inter-rater reliability studies with blind evaluation by domain experts are needed.
  \item \textbf{Fragment quality dependency.} GHOSTY COLLIDER's output quality depends heavily on fragment diversity and Ghost Extraction depth. Ghost Extraction requires practiced judgment that may vary significantly across practitioners.
  \item \textbf{AI-assistance confound.} While the protocols are designed for human-only use, the case studies were conducted with AI assistance. The protocols' effectiveness without AI has not been separately evaluated.
  \item \textbf{Absence of longitudinal and downstream validation.} Case studies demonstrate outputs but do not track whether those outputs led to successful real-world implementations. For PRECOG specifically, the Prediction Feedback Loop has not been formally exercised.
  \item \textbf{Sample size for failure rate estimation.} The batch experiment (N=8) provides an initial failure rate estimate of 12.5\%, but larger samples across diverse operators are needed for reliable confidence intervals.
\end{enumerate}
\subsection{Future Work}
\label{sec:future}
\begin{enumerate}
  \item \textbf{Multi-user controlled studies.} Pre-registered experiments with practitioners across domains and skill levels, comparing protocol-assisted ideation against unstructured brainstorming, with blind evaluation by independent domain experts.
  \item \textbf{Computational operationalization of structural novelty.} Automated metrics using semantic similarity against domain corpora, patent database novelty search, and embedding-space distance measurements.
  \item \textbf{Retroactive validation for PRECOG.} Apply PRECOG to historical events with known outcomes (e.g., the 2023 generative AI surge) to assess predictive accuracy retrospectively.
  \item \textbf{Longitudinal tracking.} Monitor PRECOG predictions against market developments. Track GHOSTY COLLIDER visions through to implementation outcomes.
  \item \textbf{Domain-specific adaptations.} Investigation of domain-specific optimizations for Ghost Extraction heuristics.
  \item \textbf{Human-only evaluation.} Test the protocols in a paper-and-pen setting without AI assistance to establish baseline effectiveness independent of AI capabilities.
\end{enumerate}
\section{Conclusion}
\label{sec:conclusion}
We have presented GHOSTY COLLIDER and PRECOG PROTOCOL as executable bridges between descriptive creativity/foresight theories and prescriptive practice. By formalizing the cognitive operations identified by Koestler, de Bono, Finke et al., Hatchuel \& Weil, Gentner, Gick \& Holyoak, Ansoff, and Wack into step-by-step protocols with explicit quality criteria and failure modes, we aim to make creative emergence and strategic foresight accessible, repeatable, and improvable.
The five case studies provide preliminary evidence of domain generality across financial analysis, business strategy, technology forecasting, music production, and video production. The batch experiment (N=8, success rate 87.5\%) provides initial quantification of protocol reliability, identifying fragment diversity as the critical success factor. The controlled comparisons suggest that the protocols produce structurally distinct outputs --- not merely better versions of conventional ideas, but qualitatively different directions that standard methods do not reach. The blind evaluation (74/80 vs.\ 49/80) confirmed the direction of author assessments while revealing a potentially larger quality gap than the author estimated.
In the context of growing concern about AI-assisted ideation reducing aggregate creative diversity \citep{doshi2024}, these protocols represent one approach to structuring human-AI interaction in ways that preserve --- and potentially enhance --- the structural novelty of creative and strategic outputs. Substantial further validation is needed, particularly multi-user evaluation with human domain expert blind assessment and retroactive validation of foresight predictions.
The protocols (v2, incorporating lessons from failure case analysis) are released as open-access documents at \url{https://github.com/GhostyAI-HA/ghosty-collider} under CC BY-NC 4.0. A companion tool, MOLD BREAKER, addressing typicality bias through structured rejection and constraint, is released alongside but is not evaluated in this paper. We welcome contributions --- particularly new case studies from diverse domains and multi-user evaluations that test the protocols' generalizability beyond their author.
\bibliographystyle{apalike}

\end{document}